# Imaging toroidic phase transitions in a direct-kagome artificial spin ice

Wen-Cheng Yue[1,2,3,+], Zixiong Yuan[1,2,3,+], Peiyuan Huang[1,2,3], Yizhe Sun[1,4], Tan Gao[1,2,3], Yang-Yang Lyu[1,2,3], Xuecou Tu[1,2,3], Sining Dong[1,3,4,*], Liang He[1,4], Ying Dong[5], Xun Cao[1], Lin Kang[1,2,3], Huabing Wang[1,2,3,*], Peiheng Wu[1,2,3], Cristiano Nisoli[6,*], and Yong-Lei Wang[1,2,3,4,*]

[1] School of Electronic Science and Engineering, Nanjing University, Nanjing, 210023, China
[2] Purple Mountain Laboratories, Nanjing, 211111, China
[3] Research Institute of Superconductor Electronics, Nanjing University, Nanjing 210023, China
[4] State Key Laboratory of Spintronics Devices and Technologies, Nanjing University, Nanjing 210023, China
[5] Research Center for Quantum Sensing, Zhejiang Lab, Hangzhou, Zhejiang, 311121, China
[6] Theoretical Division and Center for Nonlinear Studies, Los Alamos National Laboratory, Los Alamos, New Mexico 87545, USA

+ Authors contribute equally

* Correspondence to: sndong@nju.edu.cn; hbwang@nju.edu.cn; cristiano@lanl.gov; yongleiwang@nju.edu.cn

**Abstract:** Ferrotoroidicity, predicted as the fourth form of primary ferroic order, breaks both space and time inversion symmetry and represents a compelling avenue for technological advancement. However, accessing ferrotoroidicity in natural materials has proven challenging, which impedes the exploration of ferrotoroidic phase transitions. Here, we overcome the limitations of natural materials by exploring ferrotoroidicity in an artificial nanomagnet system that can be characterized at the constituent level and thermally annealed at different effective temperatures. We introduce a new nanomagnet array—the first realization of a direct-kagome spin ice. This unique artificial spin ice exhibits robust toroidal moments and a quasi-degenerate ground state, leading to two distinct low-temperature toroidal phases: ferrotoroidicity and paratoroidicity. We demonstrate experimentally and numerically a phase transition between ferrotoroidicity and paratoroidicity, along with a crossover to a non-toroidal paramagnetic phase. Our quasi-degenerate artificial spin ice in a direct-kagome structure provides a new model system to investigate magnetic states and phase transitions that are inaccessible in natural materials.

## Main

Ferroic materials are widely used in modern electronics as functional materials. They show spontaneous ordering for spin, charge and/or strain. From the point of view of the space and time parity operation there are four primary ferroic orders[1,2]: 1) ferromagnetism, a spontaneous magnetization breaking time-inversion symmetry; 2) ferroelectricity, a spontaneous charge polarization breaking space-inversion symmetry; 3) ferroelasticity, a spontaneous strain preserving both symmetries; and 4) ferrotoroidicity, a spontaneous alignment of vortices of magnetic moments violating both symmetries. One expected primary thermal dynamic property of ferroic materials is that they form long-range ordered phases when cooled below a critical temperature through the ferroic phase transition. Among the four primary ferroic orders, the ferrotoroidicity is the most elusive form, because it is difficult to characterize it experimentally in natural materials[3]. This is because its order parameter, the toroidal moment of a vortex of magnetic moments, has zero net magnetization, making it hard to detect. Consequently, the ferrotoroidic phase is still largely unexplored in nature. Here, we address this challenge by introducing a purposefully designed artificial spin ice nanomagnet array, in which toroidal phases can be visualized directly in a controllable and characterizable platform. The *direct-kagome* artificial spin ice (Fig. 1a) reveals emergent toroidal moments, while its quasi-degenerate energetics leads to emergence of ferrotoroidic and paratoroidic phases.

Artificial spin ices (ASIs) are magnetic metamaterials composed of dipolarly coupled single-domain nanomagnets[4–8]. They offer an ideal research playground for directly accessing various intriguing collective phenomena and exotic properties that might not be found in natural materials. Their basic building blocks are elongated nano-bar magnets whose magnetization can be described as a binary degree of freedom, or macro Ising spins, because of their shape anisotropy. The magnetic moments of these nanomagnets can be conveniently imaged using various magnetic imaging techniques such as magnetic force microscopy[4,9] and x-ray magnetic circular dichroism[10,11], allowing direct observation of ASI's magnetic states. In this way, interesting emergent phenomena in matter can be visually displayed in real space, including geometric frustration[4,12–15], magnetic monopoles[10,16–18], and phase transitions[19–21].

Recently, the toroidization as a ferroic order parameter was demonstrated in a so-called toroidal square ASI, in which the long-range order of ferrotoroidicity was directly imaged and manipulated[22,23]. However, such realization affords only one type of toroidic phase (the ferrotoroidic crystal phase) and the investigation of a phase transition between different toroidic phases has not been accessed. Here, we will show a unique ASI design which poses three different magnetic phases, including two distinct low energy toroidic states, thereby enabling direct investigations on the ferrotoroidic phase transition.

**Direct-kagome artificial spin ice**

The collective properties and functionalities of ASIs stem from competing dipolar interactions among nanobar magnets and are intimately tied to the geometric arrangement of the nanomagnets. The honeycomb (or traditional kagome) ASI is made of nanomagnets placed on the edges of a honeycomb lattice, thus at the vertices of a kagome lattice [10,17–20,24–29]. In contrast, we place the nanomagnets on the edges of a kagome lattice, maintaining a direct kagome lattice structure in the nanomagnets' array (Fig. 1a). To differentiate our ASI from the honeycomb ASI, which is sometimes also called 'kagome' ASI, we refer to this new ASI structure as a *direct-kagome* ASI.

As shown in Fig. 1a, the direct-kagome ASI is composed of triangular plaquettes, each with three nanomagnets in a circle (highlighted by green dashed circle in Fig. 1a), which form well-defined toroidal moments of magnetically closed loops (Fig. 1b). Figure 1b shows the toroidal moments, $\pm t$, whose polarities are determined by the chiral arrangements of the nanomagnets' moments. The toroidal moments form a honeycomb lattice with a six-fold geometrical symmetry (Fig. 1a), which is the same to the symmetry of the honeycomb ASI.

Unlike the honeycomb ASI, which has tri-legged vertices, each of the vertices in the direct-kagome ASI is made of four nanomagnets highlighted by a blue rectangular dashed box in Fig. 1a. Each vertex connects two triangular plaquettes (Fig. 1c), and the energies in the vertex determine the coupling strength between the neighboring toroidal moments. Based on their energies, the sixteen vertex configurations are divided into five groups, denominated as Type I, II-α, III, II-β and IV, respectively (Fig. 1d).

The vertex designations used here are analogous to those in a square ice, except that the Type II vertices are further separated into two groups of Type II-α and II-β vertices. As in a square ice, the low energy spin configurations of Types I and II-α satisfy the 'two-in, two-out' ice rule, and when all the vertices are in the ground state of Type I configurations, the direct-kagome ASI forms a two-fold long range ordered phase of *ferrotoroidic order* (the alignment of the toroidal moments with the same polarity) as shown in Fig. 1a and/or Fig. 2a.

**Quasi-degeneracy and thermal phases**

The vertex energetics of the direct-kagome ASI differs from that of square ice because of a shearing angle of 30 degrees[30]. This shearing of vertices lifts the degeneracy of the Type II vertices[30], resulting in a reduced energy of the Type II-α vertices and a significantly enhanced energy of the Type II-β vertices (Fig. 1d). Consequently, the excitation energy of Type II-α vertices is very low, and the energy of Type II-β vertices is higher than that of the Type III vertices (Fig. 1d). As a result, the energy $E_1$ (the energy gap between Types I and II-α vertices) of the lowest excitation Type II-α vertices is much lower than the energy $E_2$ (the energy gap between Type I and Type III vertices) of the second lowest excitation Type III vertices (see Fig. 1d). In this sense, Types I and II-α vertices can be considered quasi-degenerate at an intermediate temperature range, $E_1 \ll k_B T \ll E_2$, where $k_B$ is Boltzmann constant and $T$ is temperature[31]. Quasi-degeneracy with drastically reduced energy gap was numerically demonstrated in a connected square ASI structure[31]. Recently, a topology-restricted quasi-degeneracy was realized in a sheared square colloidal ice, where the ground-state degeneracy of two-dimensional square ice was partially recovered[30]. Here, the quasi-degeneracy in our direct-kagome ASI also partially recovers the vertex ground-state degeneracy from two-fold (only Type I) to four-fold (both Type I and Type II-α).

It follows that in the intermediate temperature range $E_1 \ll k_B T \ll E_2$, the system would be dominated by a state consisting of roughly equal occupancies of Types I and II-α vertices. This leads to an emergent low-energy toroidic phase of *paratoroidicity* (Fig. 2b). Remarkably, one can thus expect a direct toroidic phase transition between ferrotoroidicity and paratoroidicity.

We have numerically tested the various phases and phase transitions. Our Monte Carlo (MC) simulation of the temperature-dependent heat capacity and entropy (Fig. 2d) of the direct-kagome ASI clearly reveals three distinct thermal phases (Fig. 2a-2c), separated by a toroidic phase transition and a crossover, as indicated by a peak and a bump in the heat capacity curve, respectively, and two steps in the entropy curve (see Fig. 2d). The low-temperature peak signals a phase transition from ferrotoroidicity to paratoroidicity due to the spontaneous symmetry breaking in the net value of toroidal moments. The bump at higher temperature signals a crossover to nontoroidicity as Type-III vertices appear, toroidal moments disappear, and the material becomes paramagnetic.

**Imaging toroidic phase transitions**

To experimentally access the various intriguing thermal phases and phase transitions in athermal ASI samples at room temperature, we designed samples with a series of lattice constants (Figs. 3a, 3b and Extended Data Fig. 1). Based on the Boltzmann distribution law, the population probability of an excitation in thermal equilibrium is roughly proportional to exp($-E_i/k_BT$), where $E_i$ is excitation energy and $T$ is temperature. Instead of increasing the temperature $T$ we can reduce the excitation energy $E_i$ to access the *effective* thermal states in a thermal equilibrium system. Here, the vertex excitation energy is dominated by the local coupling strengths, e.g., $E_1 = 2(J_2 - J_3)$ and $E_2 = J_1 + J_2 - J_3$ in our direct-kagome ASI (see Fig. 1d). Previous investigation has shown that adjusting the local coupling strengths allows to tune the effective temperature of ASI systems[32–35]. One of the most direct ways to tune the local interactions is to regulate the ASI's lattice constant. As shown in Fig. 4a, increasing the lattice constant reduces all the coupling strengths. Moreover, the ratio between $J_1$ (the coupling defines the toroidal moment) and $J_2$-$J_3$ (the coupling defines the ferrotoroidic correlation strength between neighboring toroidal moments) remains roughly constant with changing the lattice constant (inset of Fig. 4a). This implies that adjusting the lattice constant affects all the critical energies, such as $E_1 = 2(J_2 - J_3)$ and $E_2 = J_1 + J_2 - J_3$, in the same way. This enables us to tune the effective temperature for the whole system by changing the lattice constants.

We fabricated samples with a series of lattice constant $a$ (see Fig. 1a for the definition of $a$) ranging from 300 nm to 1760 nm. To prevent nanomagnets from overlapping in samples with $a < 360$ nm, we slightly shift them away from the centers of triangular plaquettes (refer to Supplemental Information). The nanomagnet size for all the samples is 220 nm × 80 nm × 20 nm, which results in a single domain magnetization along its long axis[36]. The details of sample fabrications can be found in Methods. Figures 3a and 3b show scanning electron microscopy (SEM) images of the direct-kagome ASI with $a$ = 360 nm and 720nm, respectively (see Extended Data Fig. 1 for all the samples).

To access the low energy equilibrium states, we performed high temperature annealing for all the samples (see Methods and Extended Data Fig. 2 for details). Figures 3c and 3d show the magnetic force microscopy (MFM) images of the samples corresponding to Figs. 3a and 3b, respectively. Extended Data Fig. 3 displays MFM images of all the samples in large areas, from which we extract the microscopic spin/vertex (Figs. 3e-3i) and toroidal moment (Figs. 3j-3n) configurations, respectively (see Extended Data Figs. 4 and 5 for more data). For small lattice constants, all the triangular plaquettes form toroidal moments of closed loops, as shown by red and blue plaquettes in Figs. 3j-3l. A single domain of ferrotoroidicity is observed (Fig. 3j). As the lattice constant increases, the domain size becomes smaller and the distribution of toroidal moments become disordered (Figs. 3k-3m). When further increasing lattice constant the toroidal moments are gradually broken, as shown by the white plaquettes in Fig. 3n.

In Fig. 4b, we obtained the vertex and toroidal moment populations as a function of lattice constants. From Fig. 4a, we can extract the coupling energies $J_2$ for the samples with various lattice constant $a$. Then we plot the vertex and toroidal moment populations as a function of $1/J_2$, as shown by open dots in Fig. 4c. We also plot the temperature-dependent statistical results obtained from Monte Carlo simulations (solid lines in Fig. 4c). The nearly perfect matching between lattice constant (or coupling

strength)-dependent experiments and temperature-dependent simulations validates the accessing of thermal dynamic properties in the direct-kagome ASI by tuning its lattice constant.

One remarkable result is that the population of the Type II-α vertices increases with (effective) temperature at low temperatures and then decreases with a further increase in temperature (red in Fig. 4c). At the low temperature region, where the number of Type II-α vertices increases while that of the Type I vertices decreases. In this region, the population of the toroidal moments remains at 100%. This effect originates from the effective quasi-degeneracy of Type I and II-α vertices when gradually increasing the temperature to intermediate values. The plots show a toroidal phase transition from ferrotoroidicity to paratoroidicity. When the (effective) temperature increases further, Types III, II-β and IV vertices emerge gradually, and the population of the toroidal moments decreases. At high enough temperatures, all types of vertices and the toroidal moments saturate at their configurational (random) population rate, as the system enters a completely disordered paramagnetic state.

To quantitatively describe the three thermal phases and phase transitions in the direct-kagome ASI, we define two order parameters $\psi$ and $\varphi$ for toroidicity and ferrotoroidicity, respectively. First, we define the vorticity as $v = \pm 3, \pm 1$ for the triangular plaquettes (see the inset of Fig. 4d), where $v = \pm 3$ indicates well-defined toroidal moments with fully closed magnetic loops, while $v = \pm 1$ denotes the breaking of toroidal moments. Then the order parameters are defined as:

$$\psi = \frac{4}{3}[\psi_0 - \frac{1}{4}], \text{ where } \psi_0 = \frac{1}{2N}\sum(|v| - 1)$$

$$\varphi = \frac{1}{3N}|\sum v|,$$

where $N$ is the total number of the triangular loop units. Here, $\psi_0$ quantifies the population of well-defined toroidal moments, and the order parameter $\psi$, normalized to the range [0, 1], measures the local onset of toroidicity.

The second order parameter, $\varphi$, measures the *symmetry breaking* due to the long-range ordering of toroidal moments. It is zero if domains of opposite toroidal orientation are equally represented. The order parameters extracted from experiments (open dots) and MC simulations (solid lines) are plotted in Fig. 4d. At the lowest (effective) temperature (smallest lattice constant), both order parameters are equal to one, revealing a single domain of ferrotoroidicity. However, at an intermediate temperature, $\varphi$ decreases, while $\psi$ remains close to one, when $\varphi$ reaches zero. The system remains toroidal, but the emergence of Type II-α vertices (see Fig. 4c) disrupts the toroidal symmetry breaking, leading to paratoroidicity. Finally, at higher temperatures, $\psi$ begins decreasing and approaches to zero, as shown in Fig. 4d. This signals the system's crossover into the non-toroidal paramagnetic phase, corresponding to the appearance of high energy vertices.

We can resort to a more sensitive characterization of ferrotoroidic ordering and the associated phase transitions, by plotting the magnetic structure factor (MSF) maps for spin configurations and toroidal moment configurations in Figs. 5a-5f and 5g-5l, respectively (see Methods). Extended Data Figs. 6 and 7 show the MSF maps for all the samples. For the samples with small lattice constants (low effective temperatures), clear and sharp Bragg peaks are observed in MSF maps of both spins (Figs. 5a and 5b) and toroidal moments (Figs. 5g and 5h), which exactly match the theoretical prediction of ferrotoroidicity (Figs. 5m and 5p). This unambiguously reveals the ordering of both spin and toroidal moments required for ferrotoroidicity. One interesting feature is that the MSF maps contain both strong and weak Bragg spots, which are typical features of the ferrotoroidic ordering of direct-kagome ASI (as illustrated in Extended Data Figs. 8). With the lattice constant (or effective temperature) increasing, the spin MSF map is gradually becoming diffusive but still featured (from Figs. 5a to 5d). The featured pattern of the spin MSF represents the local correlations of the nanomagnets within each toroidal moment (the satisfying of all the $J_1$ couplings). However, the diffusive MSF pattern indicates no long-range spin ordering. On the other hand, the toroidal moment MSF map is gradually changing to featureless patterns (from Figs. 5g to 5j), suggesting that there are no correlations among the toroidal moments. These are consistent with the prediction of ferrotoroidic phase transition from ferrotoroidicity to paratoroidicity in the direct-kagome ASI (Figs. 5n and 5q). At very large lattice constants (high effective temperatures), both structure factor maps (Figs. 5f and 5l) are featureless, implying that the system transitions into a paramagnetic state (Figs. 5o and 5r).

**Conclusions**

The direct-kagome ASI structure offers an ideal platform to explore toroidal magnetism and its transitions. The quasi-degeneracy of the direct-kagome ASI, stemming from the unique structure of the Kagome lattice, leads to multiple phases. Very large excitation energies would suppress the quasi-degenerate state, thereby limiting the richness of phase diagram (see Extended Data Fig. 9). As an engineerable platform, its excitation energies can be fine-tuned through structure modifications[34,37–39], potentially leading to even more interesting phenomena. Fabricating thermally activated direct-kagome samples with ultra-thin magnetic layers would enable the study of dynamic properties[11]. Exploring connected counterparts of direct-kagome ice could unveil interesting transport effects[40] and introduce additional energy tuning parameters[31,33,35,41]. Fascinating phenomena in direct-kagome ices are also anticipated in other ice platforms, such as qubit ices[42–44], superconducting flux-quantum ices[45–48] and colloidal ices[30,49]. The direct-kagome ice could pave the way for innovative applications, including reconfigurable hybrid devices[48,50,51], programmable magnonics[52,53], and advanced computational technologies[54,55]. The violation of both space and time inversion symmetries by ferrotoroidicity not only facilitates magnetoelectric responses but also hints at potential functionalities in heterosystems. For instance, leveraging ferrotoroidicity in superconductor heterostructures might enable the creation and control of superconducting diode effects, which require simultaneous space-inversion and time-reversal symmetry breaking[56].

**Figures**

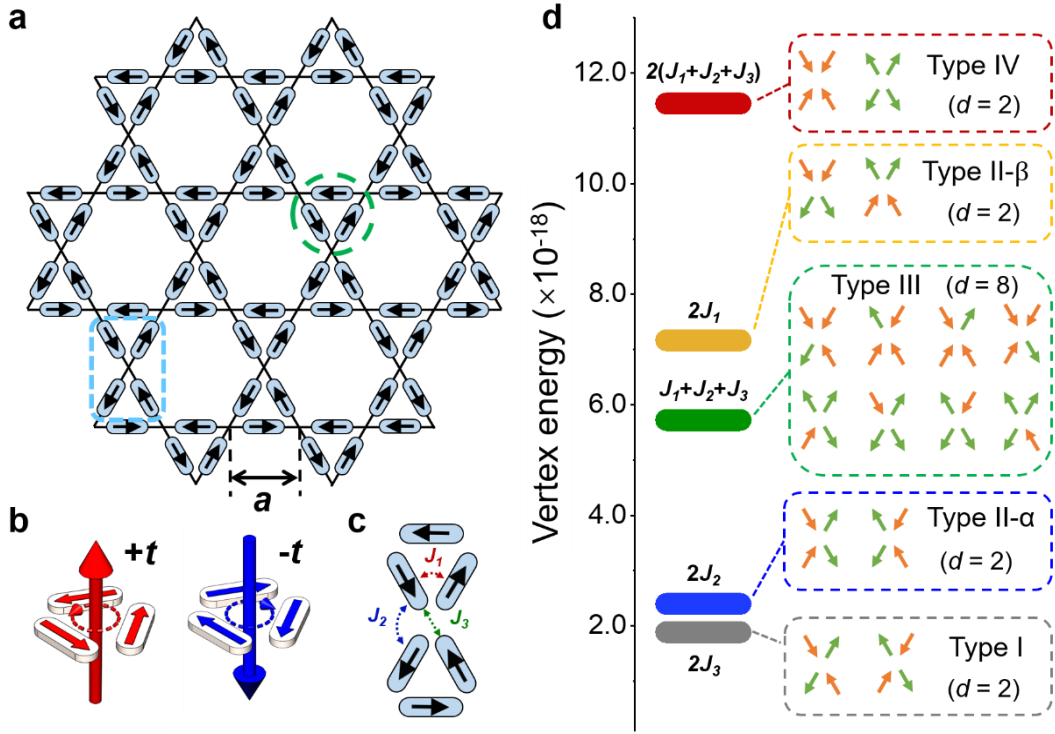

**Fig. 1 | Quasi-degenerate direct-kagome ASI with well-defined toroidal moments.**
**a**, Geometric structure of the direct-kagome ASI. A triangular plaquette and a four-leg vertex are highlighted. **b**, Toroidal moments of the nanomagnets in the triangular plaquette. **c**, Three local couplings in two neighboring triangular plaquettes. **d**, Vertex classification based on energies. Values of the degeneracy $d$ are listed for each type of vertices. The tiny energy gap between Type I and Type II-α vertices induces a quasi-degenerate state. The energies are calculated from micromagnetic simulations for nanomagnets with size of 220 nm × 80 nm × 20 nm and lattice constant $a$ = 360 nm.

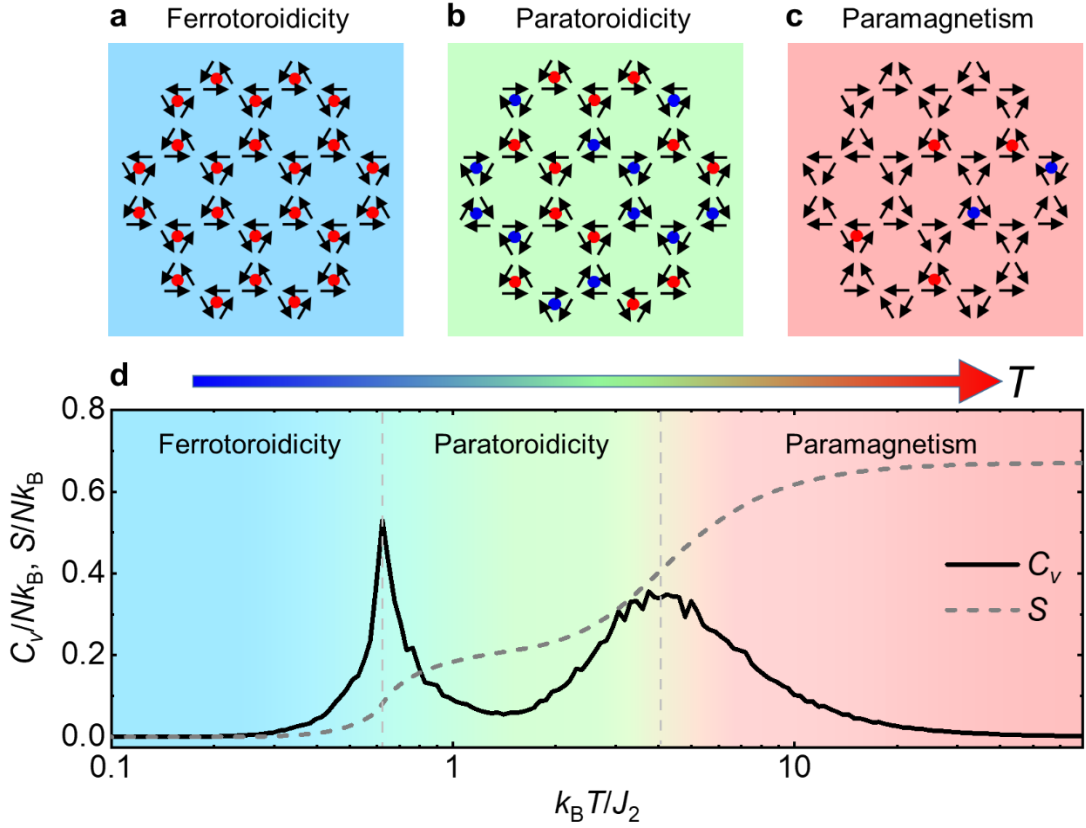

**Fig. 2 | Three thermal phases and phase transitions. a-c**, Spin and toroidal moment configurations of ferrotoroidicity (**a**), paratoroidicity (**b**), and paramagnetism (**c**). The red and blue dots represent positive (red) and negative (blue) toroidal moments. **d**, Temperature-dependent heat capacity and entropy from Monte Carlo simulation. The phase transition and the crossover among the three phases are depicted by two gray dashed vertical lines, respectively.

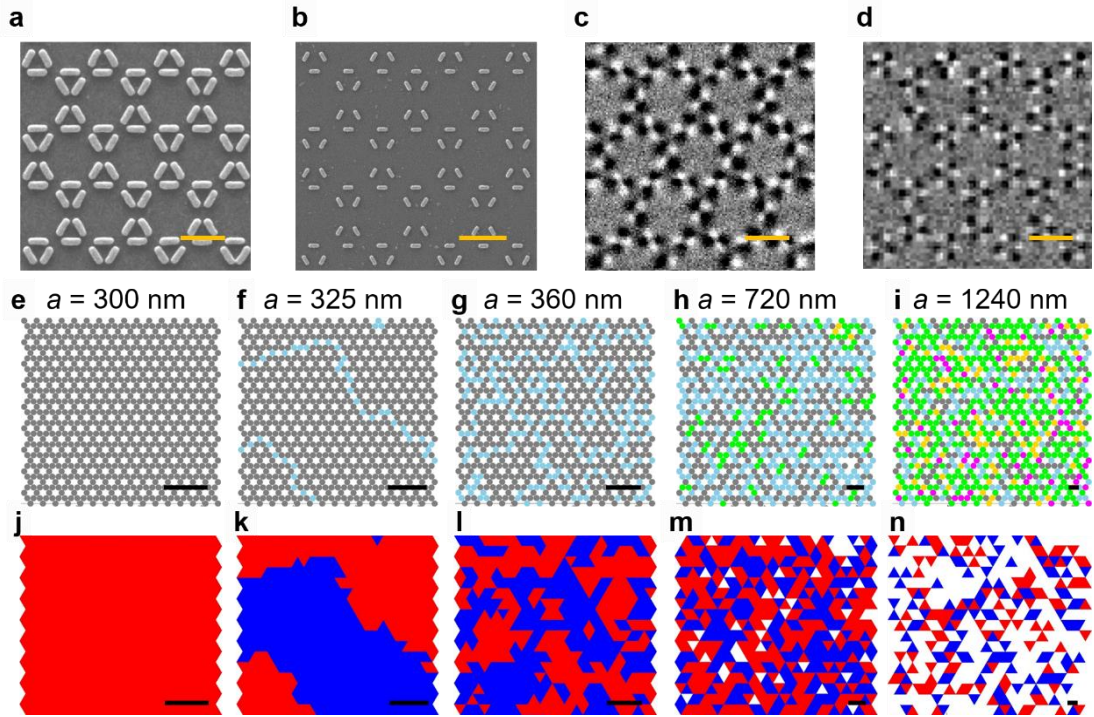

**Fig. 3 | Lattice constant dependent experiments. a** and **b**, SEM images of direct-kagome ASI with $a$ = 360 nm (**a**) and $a$ = 720 nm (**b**), respectively. **c** and **d** MFM images corresponding to (**a**) and (**b**), respectively. Scale bars, 500 nm (**a,c**) and 1 μm (**b,d**). **e-i** spin/vertex distributions extracted from MFM images with various $a$ values. Types I, II-$\alpha$, III, II-$\beta$ and IV are shown in gray, light blue, green, gold, and magenta, respectively. **j-n** toroidal moment distributions corresponding to (**e**)-(**i**), respectively. Red and blue plaquettes respectively denote positive and negative toroidal moments. Scale bars, 2 μm.

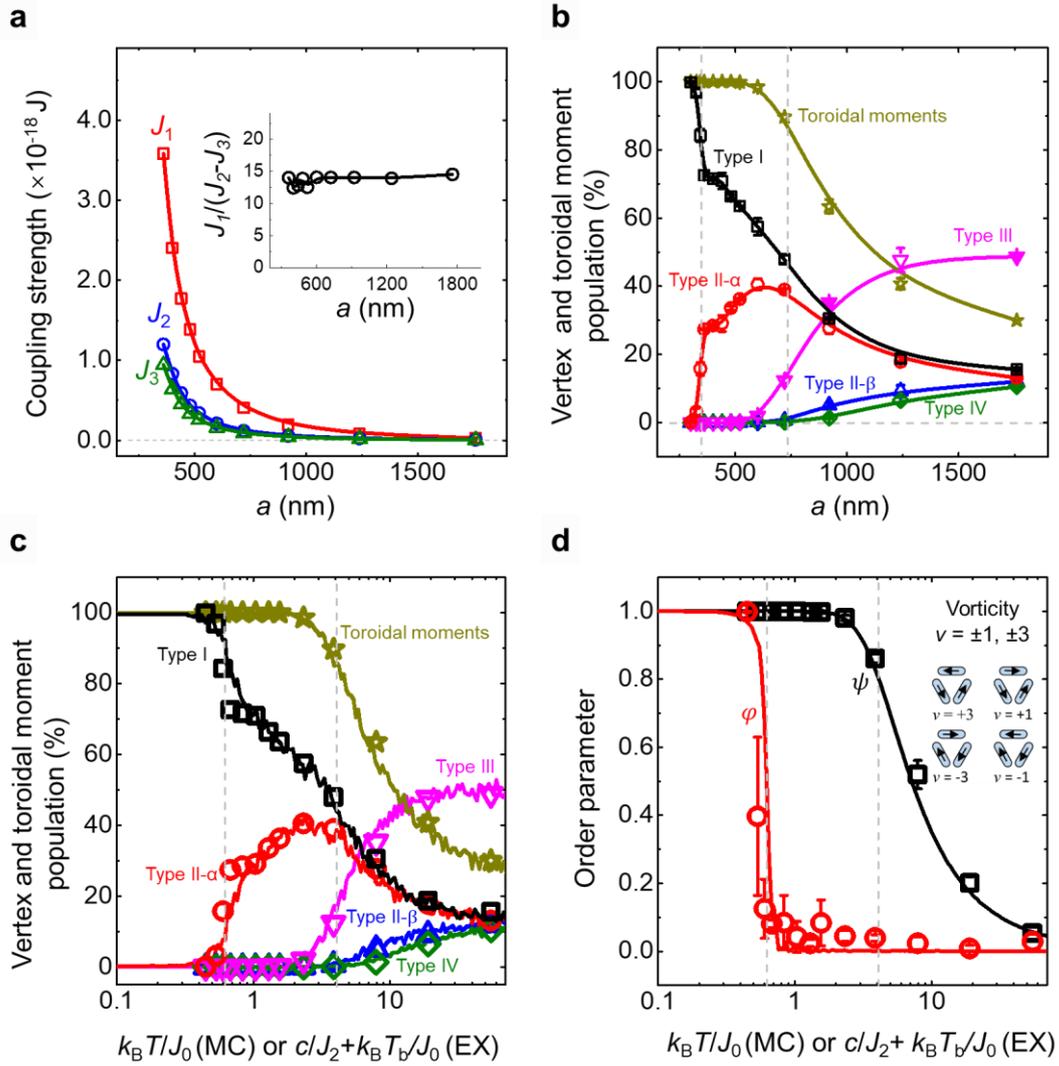

**Fig. 4 | Accessing effective thermal states by tuning lattice constant. a,** local interactions $J_1$, $J_2$ and $J_3$ (defined in Fig. 1c) as a function of lattice constant $a$, which is calculated from micromagnetic simulations. Inset: the ratio between $J_1$ and $J_2$-$J_3$. **b** statistics of the spin/vertex and toroidal moment populations extracted from MFM images. **c,** comparison of spin/vertex and toroidal moment populations between energy dependent experiments (open dots) and temperature dependent MC simulations (solid lines). Experimental (EX) results are normalized by $c/J_2+k_BT_b/J_0$, where $J_2$ is obtained from micromagnetic simulation for various lattice constants $a$ (refer to the Supplemental Information for all $J_2$ values), while the constants $c$ and $T_b$ satisfy $c = 0.35J_0$ and $k_BT_b = 0.33J_0$, with $J_0=J_2(a=360$ nm$)=1.19\times10^{-18}$ J. **d,** comparison of order parameters between experiments (open dots) and MC simulations (solid lines). The vorticity $v$ is defined in the inset, and black and red curves/dots are for the order parameters $\psi$ (toroidicity) and $\varphi$ (ferrotoroidicity), respectively. The gray dashed vertical lines in (b-d) correspond to the phase transition and crossover lines in Fig. 2d.

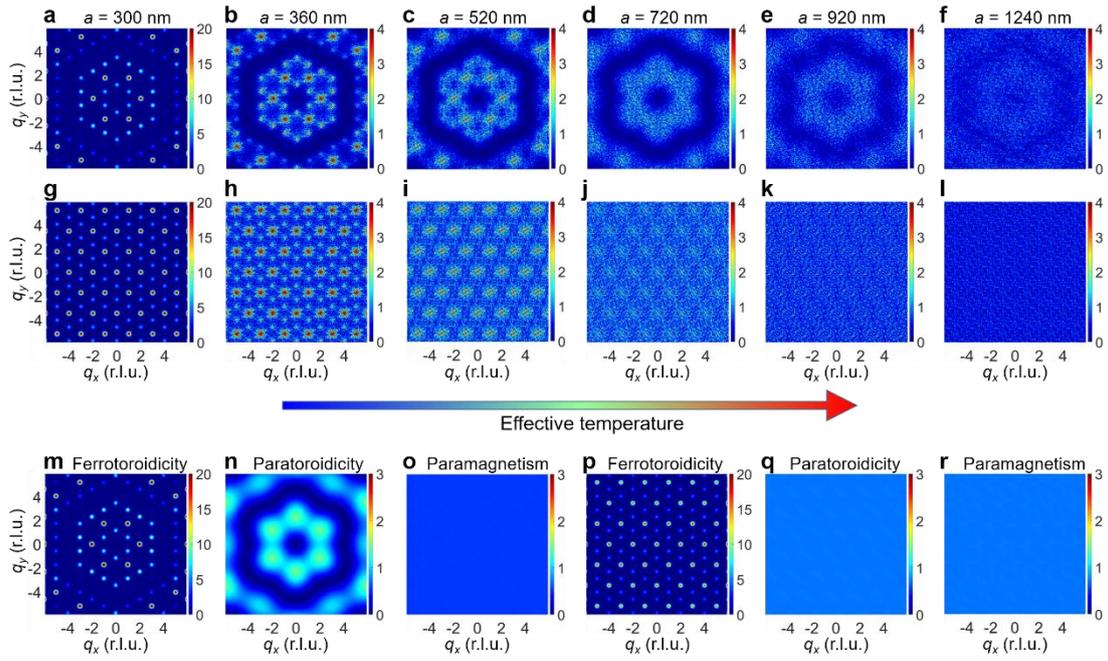

**Fig. 5 | Visualizing toroidic phase transitions from magnetic structure factors. a-l,** MSF maps of spins (**a-f**) and toroidal moments (**g-l**) deduced from experimental images for samples with various lattice constants. **m-r,** theoretical MSF maps of spins (**m-o**) and toroidal moments (**p-r**) from three ideal phases.

**Methods**

**Sample fabrication.** The ASIs' nanomagnet arrays were fabricated on a silicon substrate with a 200 nm silicon nitride layer. The nanomagnet arrays were patterned using e-beam lithography using a bilayer electron-beam (e-beam) resist of PMMA 495 (100 nm) and PMMA 950 (80 nm), followed by e-beam evaporation of 20 nm thick permalloy ($Ni_{0.8}Fe_{0.2}$) at a deposition rate of 0.3Å/s. The lift-off process was conducted by sonication in n-methyl-pyrrolidone and acetone. Detailed sample dimensions can be found in Supplemental Information.

**Thermal annealing.** The annealing process was conducted in a high vacuum chamber with a base pressure of $10^{-8}$ Torr. The sample was placed on an annealing stage covered by mu-metal shielding to prevent the influence of magnetic fields. The sample stage was heated using a laser with a beam spot diameter of 1 cm and an adjustable output power up to 100 W in continuous mode. The sample was gradually heated from room temperature to 550 °C in 60 minutes and held at this temperature for 15 min before being cooled at 0.3 °C per min to 100 °C (refer to Extended Data Fig. 2a).

**Micromagnetic simulations.** The micromagnetic simulations were performed using the *Mumax³* software package[57,58], with the following material parameters for permalloy: the exchange constant of $1.3 \times 10^{-11}$ J/m, the saturation magnetization of $8.6 \times 10^5$ A/m, and the Gilbert damping of 0.01. The mesh size is $2 \times 2 \times 2$ nm³.

**Monte Carlo simulations.** The Monte Carlo simulations were conducted by using a thermal annealing protocol. The simulations were performed on 31×31 Kagome lattice sites with free boundary conditions. The Hamiltonian is defined as $H = -\sum_{\langle i,j \rangle} J_{ij} S_i S_j$, where $S_i$ and $S_j$ are Ising variables on the sites $i$ and $j$, and $J_{ij}$ is the coupling strength between the two sites. We only consider the interactions of the nearest neighbors, $J_1$, $J_2$ and $J_3$, which are defined in Fig. 1c and are determined from micromagnetic simulations. We used the values of $J_1$, $J_2$ and $J_3$ for the sample with a lattice constant of $a = 360$ nm in our simulation. The thermal annealing protocol starts from a high temperature $k_BT/J_0=10$. Each temperature update is set to 96% of the previous temperature value. Each temperature update includes 3000 simulation steps. In each step, 1500 single-spin flip calculations are performed, followed by 750 loop update calculations. The output of each step serves as the input for the next step. The specific heat and entropy are calculated by following the reference[59].

**Magnetic structure factor.** The spin MSF was calculated by following the reference[34]. The toroidal moment MSF is calculated in the same way as the magnetic charge MSF in reference[34]. We define the vorticity $v = \pm 3, \pm 1$ (see the inset of Fig. 4d) to be in the **Z** direction, perpendicular to the sample plane. For each triangular plaquette, we consider the $\vec{v}$ at the center of the plaquette. For every vector $\boldsymbol{q} = (q_x, q_y)$ the intensity $I(\boldsymbol{q})$ is given by

$$I(\vec{q}) = \frac{1}{N} \sum_{(i,j=1)}^{N} \vec{v}_i \cdot \vec{v}_j \exp(i\vec{q} \cdot (\vec{r}_i - \vec{r}_j))$$

where $N$ stands for the number of the triangular plaquettes.

In order to simplify the calculation, the equation is divided into two parts, each containing site $i$ and $j$ respectively, and is rewritten as

$$I(\vec{q}) = \frac{1}{N} (\sum_{i=1}^{N} \vec{v}_i \exp(i\vec{q}\vec{r}_i)) \cdot (\sum_{j=1}^{N} \vec{v}_j \exp(-i\vec{q}\vec{r}_j))$$

The above equation can also be written as

$$I(\vec{q}) = \frac{1}{N} (\sum_{i=1}^{N} \vec{v}_i \cos(\vec{q} \cdot \vec{r}_i) + i\sum_{i=1}^{N} \vec{v}_i \sin(\vec{q} \cdot \vec{r}_i)) \cdot (\sum_{j=1}^{N} \vec{v}_j \cos(\vec{q} \cdot \vec{r}_j) - i\sum_{j=1}^{N} \vec{v}_j \sin(\vec{q} \cdot \vec{r}_j))$$

Then we obtain the equation

$$I(\vec{q}) = \frac{1}{N}(\vec{A} + i\vec{B}) \cdot (\vec{A} - i\vec{B}) = \frac{1}{N}(\vec{A}^2 + \vec{B}^2)$$

where $\vec{A} = \sum_{i=1}^{N} \vec{v}_i \cos(\vec{q} \cdot \vec{r}_i)$ and $\vec{B} = \sum_{i=1}^{N} \vec{v}_i \sin(\vec{q} \cdot \vec{r}_i)$. $I$ is now the quantity of toroidal moment MSF that we calculate for the interval $(q_x, q_y) = [-5.9\pi, -5.9\pi] - [5.9\pi, 5.9\pi]$ in 501×501 steps.

## Data availability

All the data supporting this study are available on the public repository https:// xxx.

## Code availability

All the codes supporting this study are available on the public repository https://xxx.

## Acknowledgements

This work is supported by the National Natural Science Foundation of China (62288101 and 62274086), the National Key R&D Program of China (2021YFA0718802), Postdoctoral Fellowship Program of CPSF, and Jiangsu Outstanding Postdoctoral Program. The work of C.N. was carried out under the auspices of the U.S. DoE through the Los Alamos National Laboratory, operated by Triad National Security, LLC (Contract No. 229892333218NCA000001) and founded by a grant from the DOE-LDRD office. Y.D. acknowledges support from the Major Scientific Research Project of Zhejiang Lab (2019MB0AD01), the Center initiated Research Project of Natural Science Foundation of Zhejiang Province (LD22F050002).

**Author contributions**
W.C.Y. and Y.L.W. conceived the project. W.C.Y., Z.Y., X.T., L.H., and L. K. fabricated the samples. W.C.Y., Z.Y., and Y.S. conducted thermal annealing. W.C.Y. and Z.Y. conducted MFM imaging. Z.Y. and C.N. performed the MC simulations and the theoretical analysis. W.C.Y. and P.H. performed the micromagnetic simulations. W.C.Y, Z.Y. and Y.Y.L. performed the statistical analysis. Z.Y. calculated the MSF. W.C.Y., Z.Y., Y.D., S.D., C.N. and Y.L.W. analysis and interpreted the data. W.C.Y., Z.Y., S.D., H.W., C.N. and Y.L.W. wrote and edited the manuscript with input from all other authors. X.C., H.W., P.W., C.N., and Y.L.W. supervised the project.

**Competing interests**
The authors declare no competing interest.

**Additional information**
Supplementary information
The online version contains supplementary material available at https://xxx.

**Correspondence and requests for materials** should be addressed to Sining Dong, Huabing Wang, Cristiano Nisoli or Yong-Lei Wang.

**Peer review information**

**Reprints and permissions information** is available at http://www.nature.com/reprints.

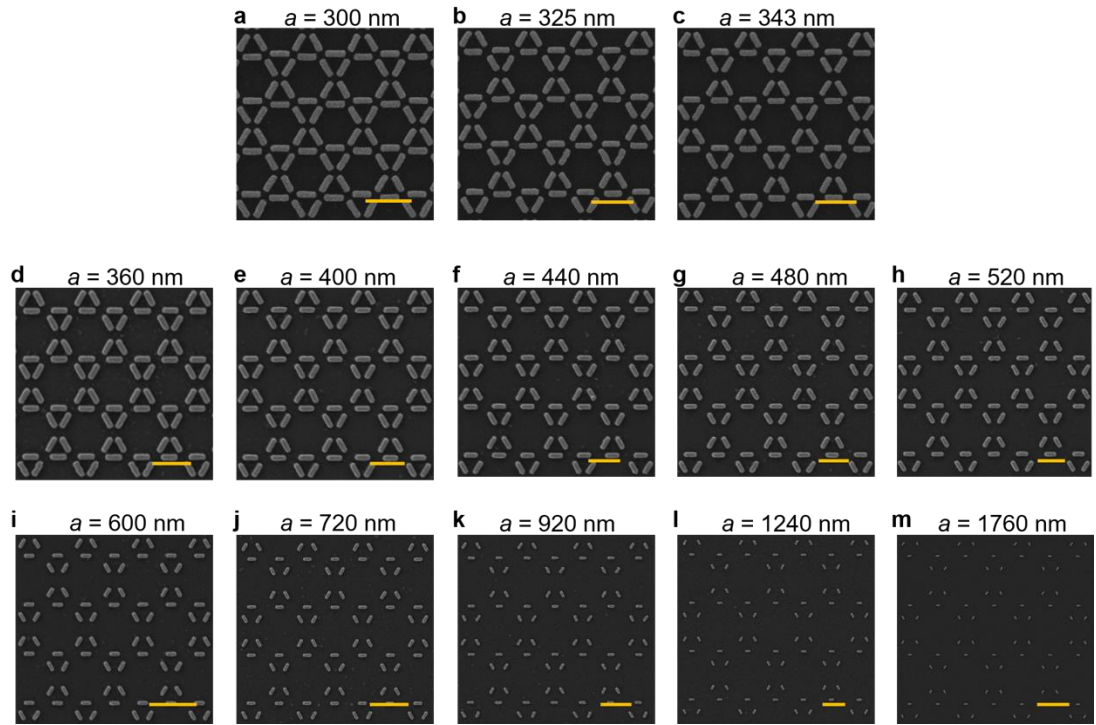

**Extended Data Fig. 1 | SEM images of samples with various lattice constants.** Scale bar, 500 nm (**a-h**), 1 μm (**i-l**) and 2 μm (**m**).

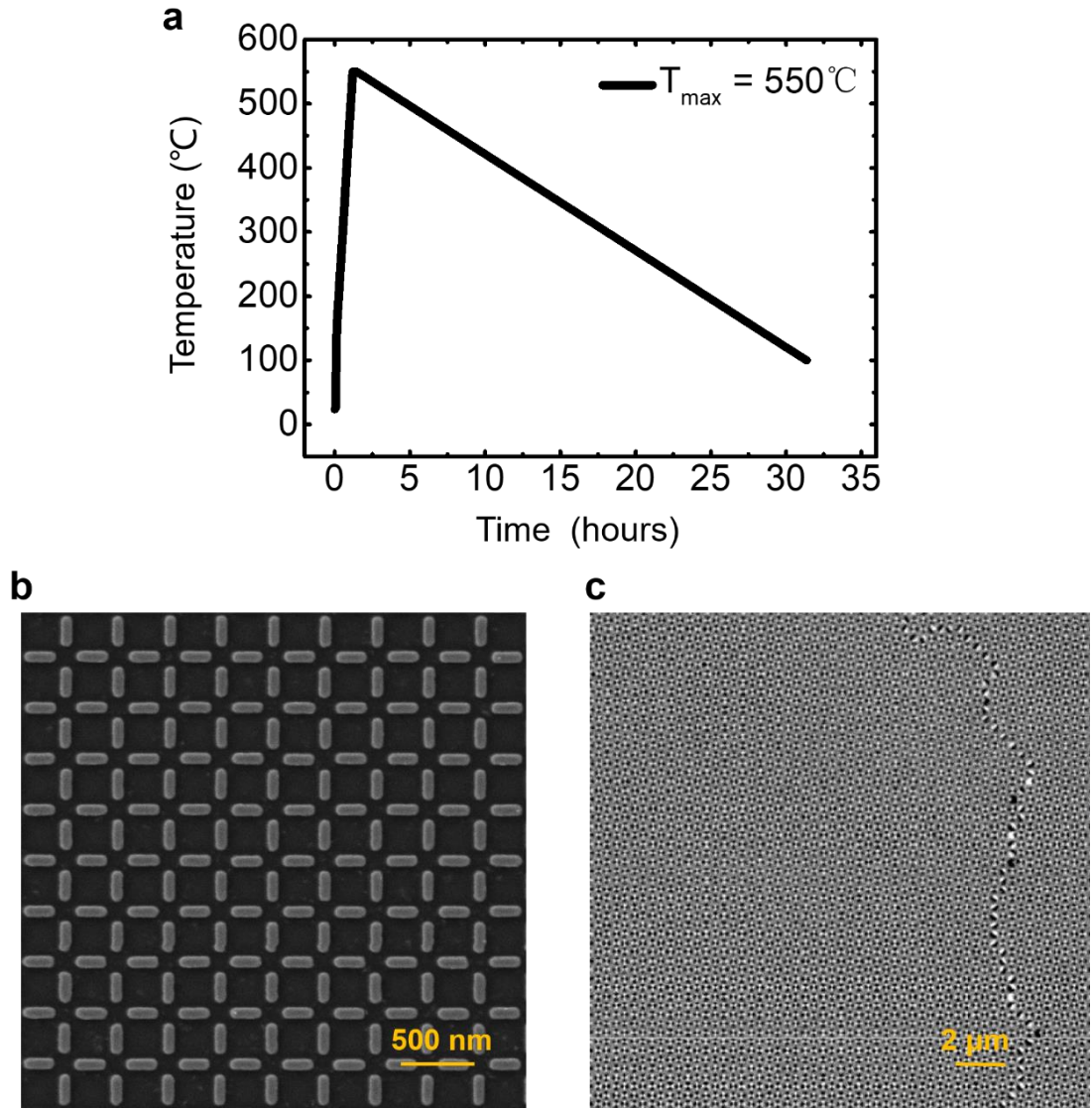

**Extended Data Fig. 2 | The annealing process. a** The sample was heated from room temperature to an annealing temperature of 550 °C in 60 minutes, held for 15 min, then cooled from 550 °C to 100 °C with 0.3 °C min$^{-1}$. **b** and **c** SEM (**b**) and MFM (**c**) images of a square ASI sample, which served as a reference and was annealed on the same substrate of the direct-kagome ASIs. The nanomagnet size is the same with that of the direct-kagome ASI. The lattice constant of the square ASI is 360 nm. The nearly perfect ground state proves the effectiveness of our annealing.

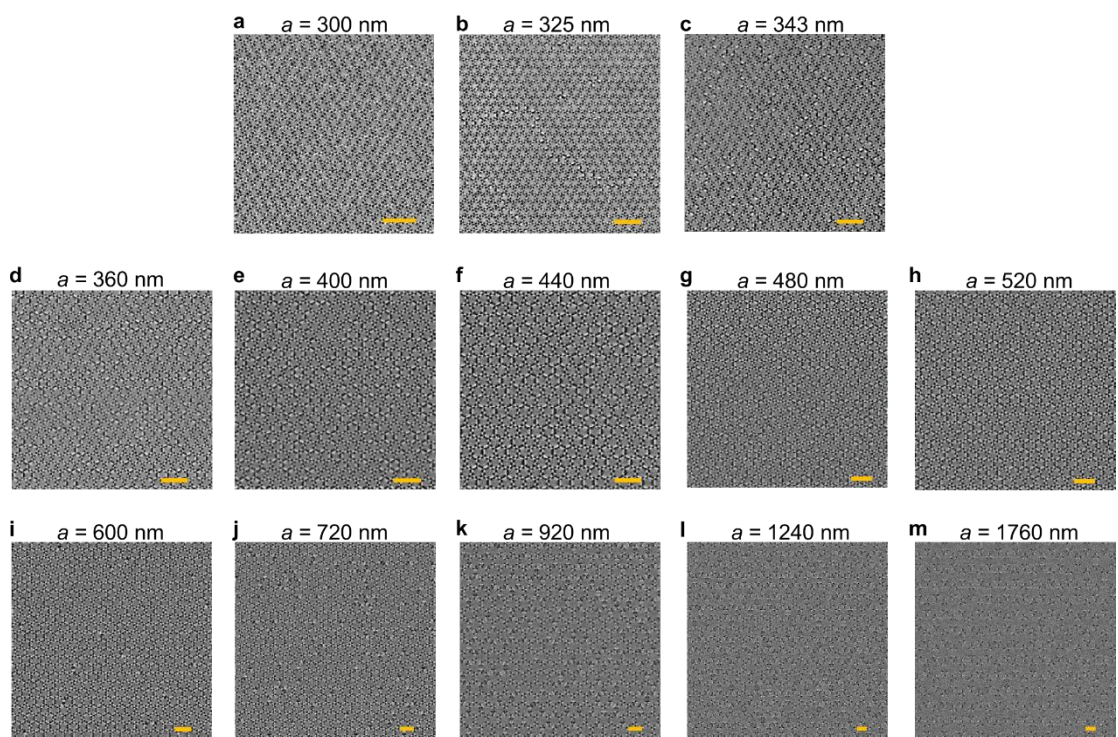

**Extended Data Fig. 3 | MFM images with various lattice constants.** The MFM imaging was conducted after a sample annealing process shown in Extended Data Fig. 1. Scale bar, 2 μm.

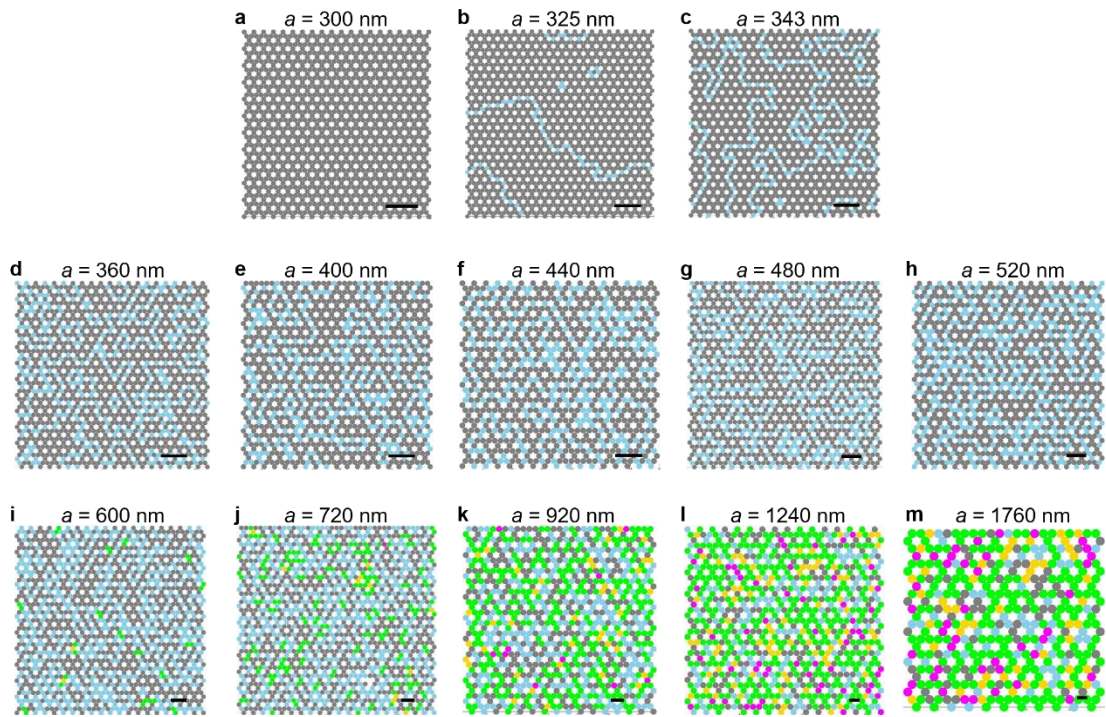

**Extended Data Fig. 4 | Vertex distributions extracted from Extended Data Fig. 3.** Vertices of Types I, II-α, III, II-β, and IV are shown in gray, light blue, green, gold and magenta, respectively. Scale bar, 2 μm.

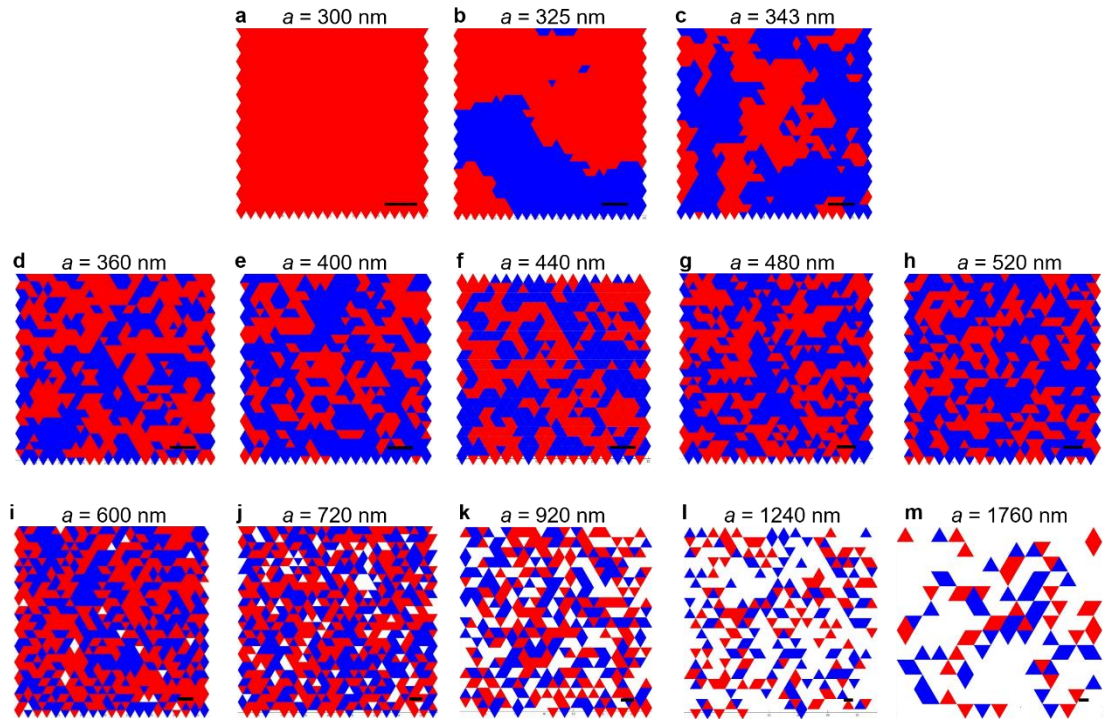

**Extended Data Fig. 5 | Toroidal moment distributions corresponding to vertex distributions in Extended Data Fig. 4.** Red and blue denote positive and negative toroidal moments, respectively. Scale bar, 2 μm.

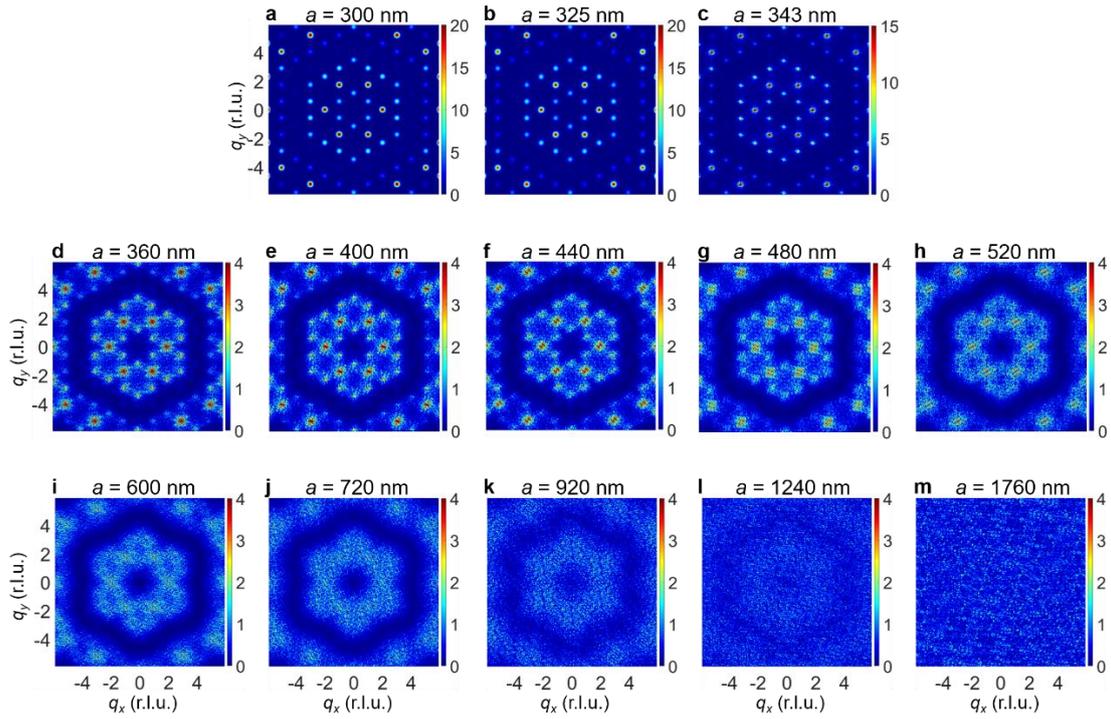

**Extended Data Fig. 6 | Spin MSF maps for all the samples. a-m** Spin MSF maps for the samples with various lattice constants ranging from 300 nm to 1760 nm, respectively.

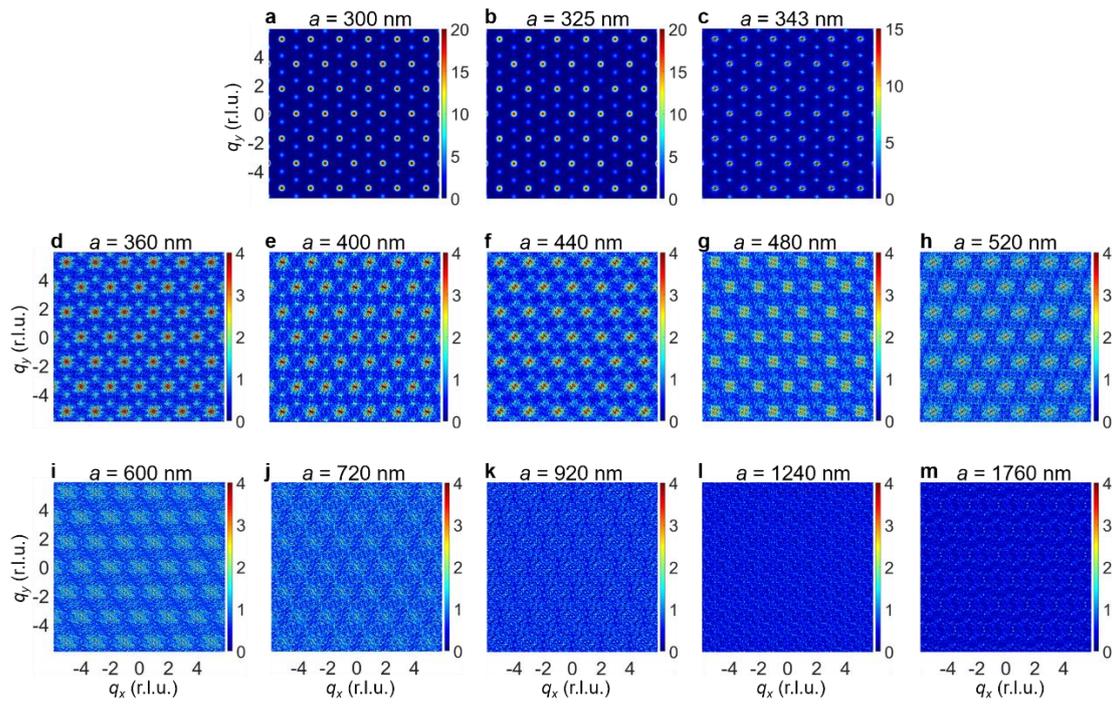

**Extended Data Fig. 7 | Toroidal moment MSF maps for all the samples. a-m** Toroidal moment MSF maps corresponding to the spin MSF maps in Extended Data Fig. 6, respectively.

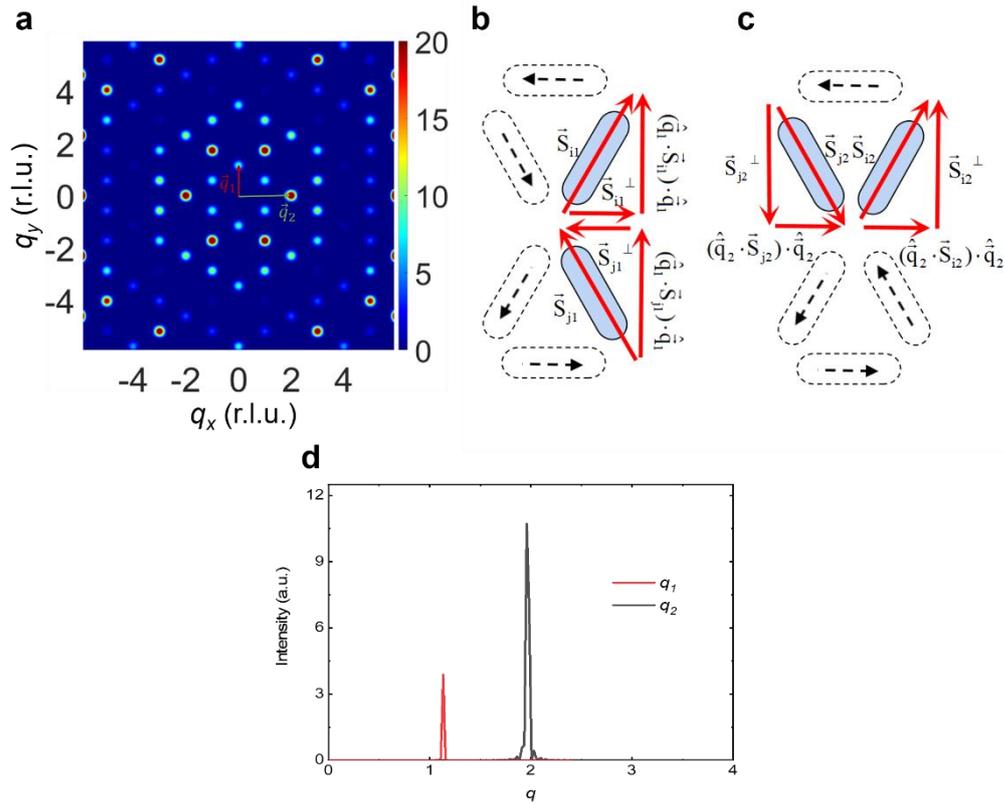

**Extended Data Fig. 8 | Strong and weak Bragg peaks in spin MSF of the direct-kagome ASI. a,** The spin MSF map of ideal ferrotoroidic ordering. A weak Bragg peak and a strong Bragg peak are marked by vectors $\vec{q}_1$ and $\vec{q}_2$, respectively. **b,** the $\vec{q}_1$ peak originates from the spin scattering between neighboring nanomagnets with an angle of 120 degrees. **c,** the $\vec{q}_2$ peak originates from the spin scattering between neighboring nanomagnets with an angle of 60 degrees. **d,** line cuts of MSF maps across $\vec{q}_1$ and $\vec{q}_2$ peaks from (a). The different spin components, $\vec{S}^\perp$, perpendicular to $\vec{q}_1$ and $\vec{q}_2$ lead to distinct intensities in the MSF for $\vec{q}_1$ and $\vec{q}_2$ (refer to Supplemental Information for a detailed derivation).

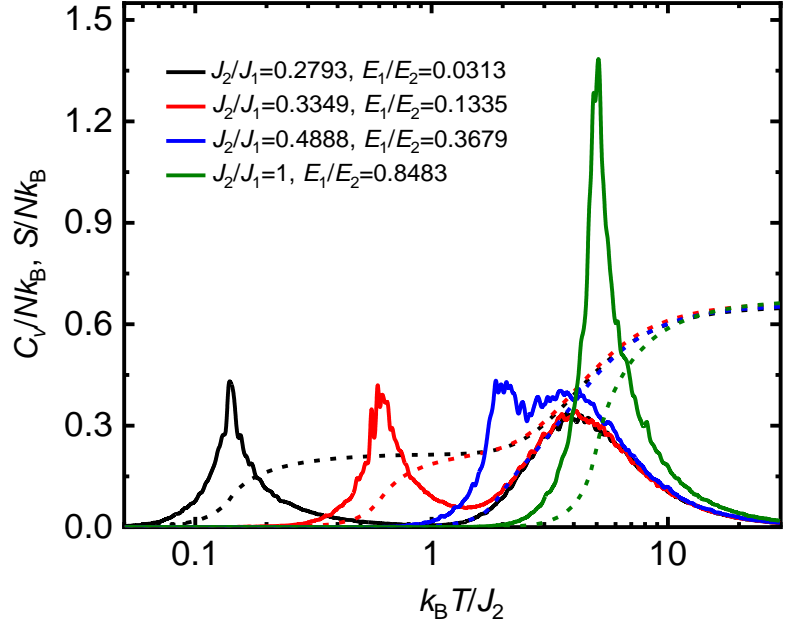

**Extended Data Fig. 9 | MC simulations of specific heat and entropy under various coupling energies.** When the lowest excitation energy $E_1$ approaching the second lowest excitation energy $E_2$, the low temperature phase transition peak is gradually merging into the high temperature crossover. Therefore, the quasidegneracy, requiring $E_1 \ll E_2$, is critical for observing the low temperature phase transition.